# Investigations of metastable $Ca_2IrO_4$ epitaxial thin-films: systematic comparison with $Sr_2IrO_4$ and $Ba_2IrO_4$

M. Souri, J. H. Gruenewald, J. Terzic, J. W. Brill, G. Cao, S. S. A. Seo

*Department of Physics and Astronomy, University of Kentucky, Lexington, KY 40506, USA*

We have synthesized thermodynamically metastable $Ca_2IrO_4$ thin-films on $YAlO_3$ (110) substrates by pulsed laser deposition. The epitaxial $Ca_2IrO_4$ thin-films are of $K_2NiF_4$-type tetragonal structure. Transport and optical spectroscopy measurements indicate that the electronic structure of the $Ca_2IrO_4$ thin-films is similar to that of $J_{eff}$ = 1/2 spin-orbit-coupled Mott insulator $Sr_2IrO_4$ and $Ba_2IrO_4$, with the exception of an increased gap energy. The gap increase is to be expected in $Ca_2IrO_4$ due to its increased octahedral rotation and tilting, which results in enhanced electron-correlation, *U*/*W*. Our results suggest that the epitaxial stabilization growth of metastable-phase thin-films can be used effectively for investigating layered iridates and various complex-oxide systems.

* Correspondence and requests for materials should be addressed to S.S.A.S. (email: a.seo@uky.edu)

**Introduction**

The spin-orbit assisted Mott state discovered in layered iridates, e.g. $Sr_2IrO_4$, provides a new platform to realize unconventional properties of condensed matter due to the unique coexistence of strong spin-orbit coupling and electron-correlation.[1] Recent studies have revealed the possibilities of novel electronic and magnetic phases in iridates such as Weyl semimetals,[2,3] and a potential high-$T_c$ superconducting state with *d*-wave gap.[4-7] However, the fundamental electronic structure of the layered iridate is still under debate; namely, the insulating gap may open due to antiferromagnetic ordering, i.e. Slater scheme,[8,9] rather than electron-correlation, i.e. Mott picture. Moreover, it is a formidable task to unveil the physics of layered iridates since only $Sr_2IrO_4$ and $Ba_2IrO_4$ (Refs. 10-13) phases are available for experimental characterizations to date.

In this article, we report the systematic changes of the structural, transport, and optical properties of layered iridates by investigating meta-stable $Ca_2IrO_4$ epitaxial thin-films. Since the Ruddlesden-Popper (*R-P*) phase of $Ca_2IrO_4$ is not thermodynamically stable, its bulk crystals do not exist in nature. However, we have successfully synthesized the *R-P* phase $Ca_2IrO_4$ thin-films (Fig. 1 (b)) from a polycrystalline hexagonal (*P62m*) $Ca_2IrO_4$ bulk crystal (Fig. 1 (a)) using an epitaxial stabilization technique.[14] The smaller ionic size of $Ca^{2+}$ compared to $Sr^{2+}$ causes increased $IrO_6$ octahedral rotation and/or tilting, hence a reduced electronic band-width (*W*). Thus, investigating $Ca_2IrO_4$ in a comparative study with $Sr_2IrO_4$ and $Ba_2IrO_4$ provides a unique opportunity to explore the layered iridate system, as it allows for the enhancement of the electronic correlation effect (*U/W*).

**Methods**

We have grown metastable $Ca_2IrO_4$ epitaxial thin-films with the $K_2NiF_4$–type crystal structure on $YAlO_3$ (110) substrates by using a custom-built pulsed laser deposition (PLD) system with *in-situ* spectroscopic monitoring techniques.[15] The laser ablation is performed on a polycrystalline hexagonal (*P62m*) $Ca_2IrO_4$ target. The powder x-ray diffraction of the target is presented in Fig. 1 (c). The samples are grown under the growth conditions of 1.2 J/cm$^2$ laser fluence (KrF excimer, $\lambda$ = 248 nm), and 700 ˚C substrate temperature. In order to avoid defects such as oxygen vacancies during the growth, we have used a laser beam imaging technique with reduced laser beam size in PLD to minimize the kinetic energy of the plume.[16] This technique is essential for the successful growth of $Ca_2IrO_4$ thin-films. A relative high oxygen partial pressure of 10 mTorr is also used to minimize oxygen vacancies. The structural properties of the epitaxial $Ca_2IrO_4$ thin-films are measured using x-ray diffractometry (Bruker D8 Advance system with Cu-Ka radiation). Transport properties are measured using a Physical Property Measurement System (Quantum Design) with conventional four-probe and Hall geometries. Optical transmission spectra ($T(\omega)$) are taken at normal incidence using a Fourier-transform infrared spectrometer in the photon energy region of 0.2–0.6 eV and a grating-type spectrophotometer in the range of 0.5–7 eV, where the substrates are transparent. The absorption spectra are calculated using $\alpha(\omega) = -\frac{1}{t} Ln(\frac{T(\omega)_{film+sub}}{T(\omega)_{sub}})$, where *t* is the thin film thickness.

**Results and Discussion**

The metastable *R-P* phase of the $Ca_2IrO_4$ thin films is verified by x-ray diffraction and reciprocal space mapping scans, which indicate that the films are stabilized by the epitaxial strain of the substrates and are of high crystalline quality. Figure 1 (d) shows the *θ-2θ* x-ray diffraction scan with the (00l) peaks of a $Ca_2IrO_4$ thin film. The full width at half maximum of the (00$\underline{12}$)-reflection rocking curve scan is 0.04°, which clearly shows good crystallinity of the film (Fig. 2 (b)). The thickness of the $Ca_2IrO_4$ thin films is ca. 6 nm. The crystal quality deteriorates considerably as we increase the thickness further, presumably due to its thermodynamically metastable nature. In x-ray reciprocal space mapping (Fig. 2 (a)), the (11$\underline{18}$) peak of the film is vertically aligned with the $YAlO_3$ substrate (332)-reflection, indicating $Ca_2IrO_4$ films are coherently strained to the substrates, i.e. $[110]_{film}$ // $[001]_{substrate}$ and $[001]_{film}$ // $[110]_{substrate}$. The lattice parameters obtained from the x-ray diffraction scans show that both in-plane (*a*) and out-of-plane (*c*) lattice parameters of $Ca_2IrO_4$ films are smaller than those of $Sr_2IrO_4$ (Ref. 17) and $Ba_2IrO_4$ (Ref. 10) (Fig. 2 (c)). At this moment, the local structural information of $Ca_2IrO_4$ films, such as octahedral rotation and tilting, is unknown and requires substantial microscopic characterizations that we plan to perform as a future study. However, by assuming the rigid Ir-O bond-length to be constant, which is a reasonable assumption, we conjecture the reduced lattice constants (from x-ray diffraction) imply that the Ir-O-Ir bond angle is reduced from 158 ° ($Sr_2IrO_4$) to ca. 140 ° ($Ca_2IrO_4$). The reduced bond angle implies a corresponding reduction in bandwidth (*W*), according to the relation between bandwidth (*W*) and the Ir-O-Ir bond angle (*θ*) described by:[18]

$$W \approx \frac{\cos\{(\pi - \theta)/2\}}{d_{Ir-O}^{3.5}} \qquad (1)$$

, where $d_{Ir-O}$ is the Ir-O bond length. This will result in an enhanced electron-correlation ($U/W$) for the $Ca_2IrO_4$ compound as compared to that of the $Sr_2IrO_4$ and $Ba_2IrO_4$ thin films.

Figure 3 (a) shows the temperature-dependent resistivity $\rho(T)$ of a $Ca_2IrO_4$ thin film, which has an insulating behavior. The room-temperature resistivity of $Ca_2IrO_4$ (ca. 170 mΩcm) is about the same as the room temperature resistivity of $Sr_2IrO_4$ (ca. 140 mΩcm) and $Ba_2IrO_4$ (ca. 130 mΩcm) deposited on $SrTiO_3$ substrates. The energy gap ($\Delta = 2E_a$) of $Ca_2IrO_4$ is calculated using an Arrhenius plot ($\rho = \rho_0 e^{\Delta/2k_BT}$, where $k_B$ is the Boltzmann constant) and compared to $Sr_2IrO_4$ (Ref. 10) and $Ba_2IrO_4$ thin films. While the Arrhenius plots of $Sr_2IrO_4$ (Ref. 10) and $Ba_2IrO_4$ show non-linear behaviors, the transport of $Ca_2IrO_4$ is quite linear over the entire measured temperature range (300 K to 90 K). An energy gap of 120 meV is extracted from its Arrhenius plot. Due to the increased $U/W$ in $Ca_2IrO_4$, we expect its gap energy to be larger than that of $Ba_2IrO_4$ and $Sr_2IrO_4$. However, the energy gap of $Ca_2IrO_4$ obtained from the room temperature transport is smaller than that of $Sr_2IrO_4$ and $Ba_2IrO_4$. This puzzling observation implies that the transport properties of layered iridates are mostly dominated by impurities or defects, and intrinsic bandgap energies should be measured using a spectroscopic technique.

Figure 3 (b) presents the optical absorption spectra ($\alpha(\omega)$) of $Ca_2IrO_4$ compared with $Sr_2IrO_4$ (Ref. 17) and $Ba_2IrO_4$ (Ref. 10) thin films. The absorption spectra are fit using a minimal set of Lorentz oscillators. The common features of strong absorption tails due to the charge-transfer transitions from O 2$p$ to Ir 5$d$ bands are above ca. 2 - 3 eV. The black solid lines in Fig. 3 (b) are the resultant fit curves using Lorentz oscillators, which match well with the experimental spectra. The three absorption peaks indicated by $\alpha$, $\beta$, and $\gamma$ are labeled consistently with previous literature.[19,20] The $\alpha$, $\beta$, and $\gamma$ absorption bands have been interpreted as the associated Ir $5d$ transitions, such as Ir-Ir intersite optical transitions.[1,19,20] Note that as the cation size — and consequently the Ir $5d$ bandwidth — increases from $Ca_2IrO_4$ to $Ba_2IrO_4$, the $\alpha$, $\beta$, and $\gamma$ peak-positions are shifted to *higher* energy. This seemingly counterintuitive peak shift has also been observed in the optical absorption spectra of strain-dependent $Sr_2IrO_4$ thin-films,[17] as the lattice strain changes from compressive to tensile directions. This observation of the peak-energy shift can provide a key to understanding the electronic structures of iridates since the spectral shape is thought to be strong experimental evidence supporting the Mott picture of this system.[1,19,20] However, we will leave it as a future study since detailed analysis requires theoretical modeling and calculations, which is beyond the scope of this article.

We note the increased optical gap energy of $Ca_2IrO_4$ thin-films as compared to that of $Sr_2IrO_4$ and $Ba_2IrO_4$. To calculate the optical energy gap, each absorption spectrum is fit using the Wood-Tauc's method[21] (Fig. 3 (c)). In this method, the strong region of the absorption edge ($\alpha > 10^4$ cm$^{-1}$) can be described by:

$$\alpha \approx \frac{(E - E_g)^\gamma}{E} \qquad (2)$$

where $E_g$ ($E$) is the optical band gap (incident photon) energy. The estimated optical gap energies using this method are $\Delta_{CIO} = 210$ meV, $\Delta_{SIO} = 150$ meV, and $\Delta_{BIO} = 110$ meV. For the exponent $\gamma$, we have obtained $\gamma = 1.5$ ($Ca_2IrO_4$), $\gamma = 3.0$ ($Sr_2IrO_4$), and $\gamma = 1.5$ ($Ba_2IrO_4$). While $\gamma = 3$ is consistent with the indirect bandgap of $Sr_2IrO_4$, $\gamma = 1.5$ values in $Ca_2IrO_4$ and $Ba_2IrO_4$ suggest direct gap, of which physical understanding will require further theoretical studies. Nevertheless, as shown in Fig. 3 (c), the optical gap energy has clearly increased for $Ca_2IrO_4$ compared to that of $Sr_2IrO_4$ and $Ba_2IrO_4$. Hence, as we decrease the ionic sizes of A-site cations in layered iridates, the Ir-O-Ir bond angle is reduced, which, in turn, increases $U/W$ and manifests itself as the observed increase in the optical bandgap energy.

Our approach of synthesizing meta-stable phase thin-films of strongly correlated systems offers a new route to understanding the physics of complex oxides. For example, the stabilization of metastable phases can provide compounds with larger effective electronic correlations than presently available by producing increased distortion and tilting in lattice. While simple octahedral distortions usually preserve inversion symmetry in the $K_2NiF_4$–type structure, the *R-P* structure of $Ca_2IrO_4$ has been proposed as a candidate material featuring a non-centrosymmetric structure due to its low symmetry.[22] This unique structure, achieved by breaking the inversion symmetry in this system, is expected to induce many interesting phase transitions such as ferroelectricity and multiferroicity. Hence, experimental studies of meta-stable phases allow us to tackle a number of intriguing problems of exotic ground states with novel properties that are theoretically suggested.

**Conclusion**

We have successfully stabilized $Ca_2IrO_4$ thin-films with the $K_2NiF_4$–type crystal structure and determined its higher optical gap energy to originate from its enhanced electron-correlation, $U/W$, with respect to its larger *A*-site cation isosymmetric compounds. The structural study confirms the good crystallinity and coherent strain state of the epitaxial $Ca_2IrO_4$ thin-films on $YAlO_3$ (110) substrates. The transport and optical spectroscopic experiments show that $Ca_2IrO_4$ thin-films have an insulating ground state similar to $Sr_2IrO_4$ and $Ba_2IrO_4$. However, the increased $IrO_6$ octahedral rotation, tilting, or distortion in $Ca_2IrO_4$ increases $U/W$, and thus its optical gap energy is larger than the gap energies of $Sr_2IrO_4$ and $Ba_2IrO_4$. This approach of metastable thin-film phases can greatly expand the number of available materials and can help to unveil the physics of strongly correlated systems.

**Acknowledgements**

We acknowledge the support of National Science Foundation grants DMR-1454200 (for thin-film synthesis and characterizations), DMR-1265162 (for target synthesis), and DMR-1262261 (for infrared spectroscopy) in addition to a grant from the Kentucky Science and Engineering Foundation as per Grant Agreement #KSEF-148-502-14-328 with the Kentucky Science and Technology Corporation.

**Author Contributions**

M.S. and S.S.A.S. synthesized the thin-film samples. M.S. carried out the x-ray diffraction, transport, and optical measurements. M.S., J.H.G., and S.S.A.S analyzed the experimental data. G.C. and J.T. have synthesized the polycrystalline target. M.S., J.H.G., and J.W.B. conducted the FT-IR experiments. M.S. and S.S.A.S. wrote the manuscript and all the authors reviewed the manuscript. S.S.A.S. initiated and supervised the project.

**Additional Information**

**Competing financial interests:** The authors declare no competing financial interests.

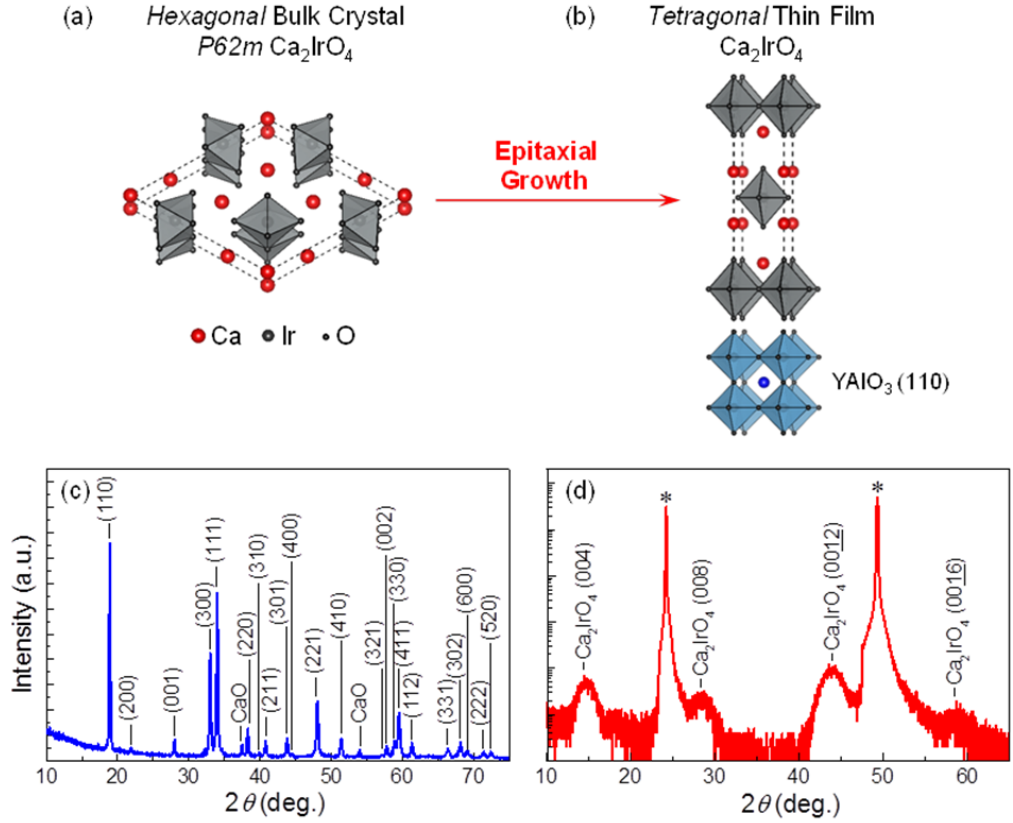

**FIG. 1.** Schematic diagram of epitaxial stabilization of tetragonal $Ca_2IrO_4$ epitaxial thin-film from (a) the bulk hexagonal phase of $Ca_2IrO_4$, i.e. a target used in the pulsed laser deposition, to (b) metastable *R-P* phase of $Ca_2IrO_4$ thin-film grown on a single crystal $YAlO_3$ (110) substrate. (c) Powder x-ray diffraction of our target material, which shows x-ray diffraction peaks from the hexagonal bulk phase of *P62m* and a couple of small CaO peaks. (d) X-ray $2\theta$-$\omega$ scan of an epitaxial $Ca_2IrO_4$ thin-film, where only the (00*l*)-diffraction peaks of $Ca_2IrO_4$ are visible. $YAlO_3$ (110) and (220) peaks are labeled with asterisk (*) symbols.

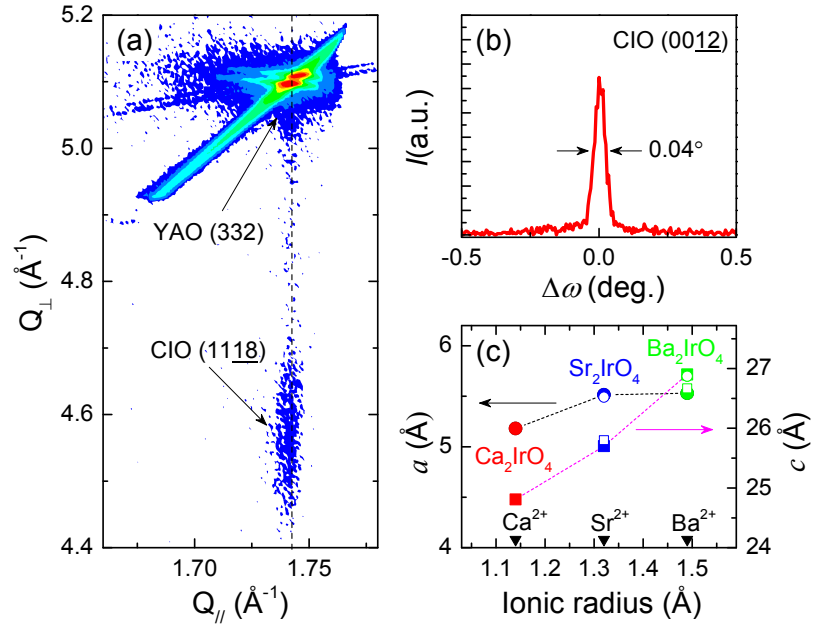

**FIG. 2.** (a) Reciprocal space map near the YAlO$_3$ (332)-reflection, which shows the Ca$_2$IrO$_4$ (11$\underline{18}$)-reflection. The vertical dashed line indicates that the Ca$_2$IrO$_4$ thin-film is coherently strained to the substrate. (b) The rocking curve scan of Ca$_2$IrO$_4$ (00$\underline{12}$)-reflection has a full-width half-maximum of 0.04°. (c) The in-plane (left axis) and out of plane (right axis) lattice parameters of Ca$_2$IrO$_4$, Sr$_2$IrO$_4$ (Ref. 17) and Ba$_2$IrO$_4$ (Ref. 10) thin films obtained from x-ray diffraction scans, as a function of A-site cation ionic radius. The solid circles and squares present the in-plane and out of plane lattice parameters, respectively. The open symbols indicate the in-plane and out of plane lattice parameters of Sr$_2$IrO$_4$ and Ba$_2$IrO$_4$ single crystals.[12,23,24]
12

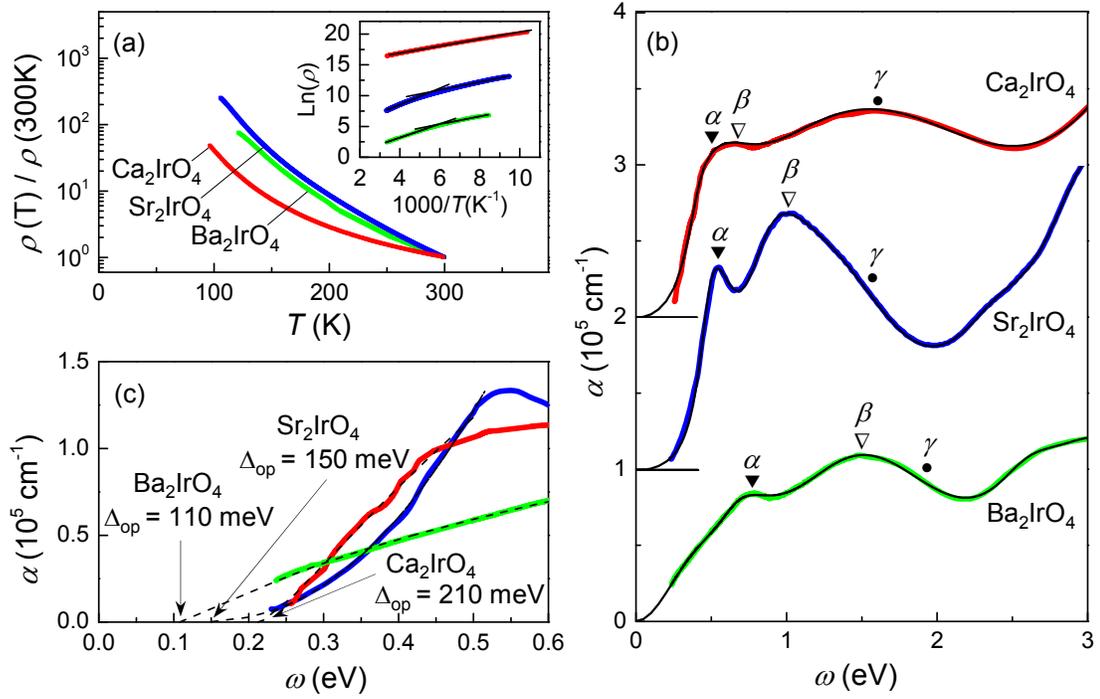

**FIG. 3.** (a) Normalized resistivity ($\rho$) versus temperature data of $Ca_2IrO_4$ (red), $Sr_2IrO_4$ (blue) and $Ba_2IrO_4$ (green) thin-films. The data of $Sr_2IrO_4$ is from Ref. 10; The inset shows the Arrhenius plot of $Ca_2IrO_4$, $Sr_2IrO_4$ and $Ba_2IrO_4$. Solid black lines present the linear fits at room temperature and low temperature. The estimated gap energies at room temperature are $\Delta_{CIO}$ = 120 meV, $\Delta_{SIO}$ = 250 meV, and $\Delta_{BIO}$ = 190 meV. The Arrhenius plots are shifted vertically for clarity. (b) Optical absorption spectra ($\alpha(\omega)$) of $Ca_2IrO_4$, $Sr_2IrO_4$ and $Ba_2IrO_4$ thin-films at room temperature. The plots are shifted vertically by $10^5$ cm$^{-1}$ for clarity. The $\alpha$, $\beta$ and $\gamma$ represent the optical transition peak energies obtained from the fit with the minimal set of the Lorentz oscillators. The solid black curves are the fit curves using Lorentz oscillators, which match well with the experimental spectra. (c) Fitted absorption spectra of $Ca_2IrO_4$, $Sr_2IrO_4$ and $Ba_2IrO_4$ at low energy using Wood-Tauc's method[21] which clearly confirm the increased energy gap from Ba$_2$IrO$_4$ to Ca$_2$IrO$_4$. The estimated optical gap energies using this method are $\Delta_{CIO}$ = 210 meV, $\Delta_{SIO}$ =150 meV, and $\Delta_{BIO}$ = 110 meV.